\documentstyle[eqsecnum,aps]{revtex}

\begin{document}
\draft
\title{Thermalized Displaced and Squeezed Number States in the Coordinate
             Representation}
\author{Wen-Fa Lu}
\address{CCAST(World Laboratory) P.O. Box 8730, Beijing, 100080,
  \\ and \\
Department of Applied Physics, Shanghai Jiao Tong University,
Shanghai 200030, China
   \thanks{mailing address,E-mail: wenfalu@online.sh.cn}   }
\date{\today}
\maketitle

\begin{abstract}
Within the framework of thermofield dynamics, the wavefunctions of the
thermalized displaced number and squeezed number states are given in the
coordinate representation. Furthermore, the time evolution of these
wavefunctions is considered by introducing a thermal coordinate representation,
and we also calculate the corresponding probability densities, average values
and variances of position coordinate, which are consistent with results in the
literature.
\end{abstract}

\section{Introduction}
\label{1}

Displaced number states and squeezed number states are generalization of
coherent states and squeezed states of a harmonic oscillator, respectively
\cite{1}. The coherent state is constructed by displacing the ground state
of the harmonic oscillator \cite{2,3}, and the squeezed state by first
squeezing the ground state and then further displacing it ( sometimes, by
first displacing and then squeezing, or by only squeezing ) \cite{4}. In these
constructions, number states ( also called Fock states in quantum field theory
) of the harmonic oscillator taking the place of the ground state will
correspondingly produce the displaced number state and squeezed number state.
Thermalizing the displaced and squeezed number states, one can get the
thermalized displaced number state and squeezed number state, which will be
discussed in the present paper. Evidently, the thermalized coherent and
squeezed states are the special cases of the thermalized displaced number and
squeezed number states. All the above states are interesting and important in
physics.

As is well known, the coherent state can describe the coherent light, and its
over-completeness gives rise to the coherent-state representation which is
very useful in quantum optics, statistical physics, quantum field theory and
particle physics, etc. \cite{5,6}. This state mimics the motion of classical
particles, and hence is also used for studying the Schr\"odinger cat states
\cite{7}. For the squeezed state, not only is it a minimum uncertainty state
and similar to the classical motion, but also the quantum fluctuations in
position can be suppressed at the expense of enhanced fluctuations in momentum
\cite{8}, which is different from that of the coherent state. Therefore, the
squeezed state has important technology applications in quantum computation
and sensitive measurement \cite{9}. So far the squeezed state has been
receiving a great deal of investigations \cite{5,10}. In the same way, a
displaced number state follows the motion of a classical particle as well as
keeps its shape in the course of the motion \cite{1}, and the squeezed number
state can display the similar squeezed property to the squeezed state
( Eqs.(38)---(41) in Ref.~\cite{1} ) and hence promises hopeful applications in
optical spectroscopy, communications, molecular and solid state physics
\cite{11} ( This reference dealed with the squeezed displaced number state )
\footnote{The author thanks the referee for recommending this reference}.
Early in 1950's, the displaced number state and the squeezed number state were
proposed and studied with the help of the coordinate representation \cite{12}.
Since then, these states received a few further investigations \cite{13,14}.
Recently, Ref.~\cite{1} reviewed these investigations, and gave the most
general time-dependent wavefunctions and probability densities of them in the
coordinate representation. Particularly, in view of the expremental realization
of the optical and atomic squeezed ( not displaced ) states as well as the
number states \cite{9} (1996) \cite{15}, Nieto predicted that in the not too
distant future, it is hopeful to observe the displaced and squeezed number
states \cite{1}.

On the other hand, no thermal noises exist nowhere, and thus the influence
of the noises on the above-mentioned states has to be studied. Such a
investigation is often realized by using density matrices and master equation.
However, within the framework of thermofield dynamics \cite{16}, a
thermalizing operator acting on the states is also an important and useful way
to introduce finite temperature effects \cite{17,18}. Both the
density-matrices and the thermofield-dynamics investigations give rise to a
varieties of thermal partners of the above-mentioned states, such as the
thermalized coherent, squeezed, displaced number, and squeezed number states,
the displaced thermalized state, squeezed thermalized state, and so on. Many
properties of various thermal coherent and squeezed states have been studied by
constructing directly a state vector \cite{17,18,19} and other methods, such as
characteristic function, density operators, Glauber's P-representation of
density operator, etc. \cite{14,20,21,22,23} ( Most of Refs.~\cite{20,21} were
concerned with the coherent and squeezed thermalized states ). The connections
between these thermal states have been revealed in Ref.~\cite{18}, and Fearn
and Collett also gave the physical interpretations of these states \cite{18}.
Besides, for the thermal coherent state, Barnett and Knight discussed the
independence of the Glauber's p-representation upon the order of displacing and
thermalizing operators \cite{17}. As for the thermalized displaced number and
squeezed number states, there were few investigations of them, and just
recently the thermalized squeezed number state ( not displaced, different from
 the state in the present paper ) was considered with its characteristic
 function for analysing the influence of thermal noise on higher-order
 squeezing properties of it \cite{24}

This paper will address the wavefunctions and position probability densities
of the thermalized displaced number and squeezed number states ( Hereafter,
two of these states will also imply that the thermalized coherent and squeezed
states are their special cases). This problem hasn't, to our knowledge,
discussed in the literature, except for Ref.~\cite{20}(1993) and Ref.~\cite{22}
(1965) in which the position probability density of coherent thermal state and
squeezed thermal state ( not include number state ) was given by Glauber's
R-function and/or P-representation ). However, this problem is certainly
interesting and meaningful. The wavefunctions of the coherent, squeezed,
displaced number and squeezed number states contain all information about
these states and hence describe completely these tates. Therefore, the
wavefunctions of the corresponding thermalized non-classical states will give
the influence of finite temperature on the properties described by the
zero-temperature wavefunctions, and can provide, at least, a
quantum-mechanical intuitional understanding for us. Moreover, the coordinate
representation of their density operators can be obtained from the
finite-temperature wavefunctions and consequently these wavefunctions can
equip a coordinate-representation way for calculating the expectaion velues of
all physical observables on the thermalized non-classical states, which is
most usual way in quantum mechanics. Additionally, the position density
probability can give the probability density of magnetic component of
electromagnetic fields \cite{25} Ref.~\cite{22}(1965).

Thermofield dynamics is unique formalism of finding the wavefunctions for the
thermal non-classical states. In this paper, within the framework of
thermofield dynamics, we shall give wavefunctions of the thermalized displaced
number and squeezed number states in terms of the position coordinate,
consider their time evolution, and calculate the position probability
densities. In order to do so, we shall first derive the wavefunction of the
thermal vacuum in the coordinate representation, which was almost given in
Ref.~\cite{8}, and introduce a thermal coordinate representation in the next
section. Then the wavefunctions of the thermalized displaced number and
squeezed number states will be given in terms of the position coordinate in
Section III. Section IV will address the time evolution, the position
probability densities, the position average values and variances of these
states. We will conclude this paper at the end.

By the way, thermofield dynamics will be not introduced in this paper, and
good expositions of them can be found in Ref.~\cite{16}. Besides, although this
paper will discuss a harmonic oscillator with a mass and constant frequency,
taking the mass as unit one can get the results which are usable for a one-mode
electromagnetic field with the same frequency.

\section{Thermal Vacuum and Thermal Coordinate Representation}
\label{2}

In the fixed-time Sch\"odinger picture, for the quantum one-dimensional
oscillator 
\begin{equation}
H={\frac {1}{2m}}p^2 +{\frac {1}{2}} m\omega^2 x^2
  =(a^\dagger a+{\frac {1}{2}})\hbar\omega \;,
\end{equation}
the ground state in the coordinate representation is the wavefunction
\begin{equation}
<x|0> = ({\frac {m\omega}{\pi \hbar}})^{\frac {1}{4}}
      exp\{-{\frac {m\omega}{2\hbar}}x^2\}  \;,
\end{equation}
where $p=-i\hbar{\frac {d}{dx}}\equiv -i\hbar \partial_x$, $m$ is the mass,
$\omega$ the angular frequency, and
\begin{equation}
a={\frac {1}{\sqrt{2m\hbar\omega}}}(ip+m\omega x) \;, \;\;\;\;\;
a^\dagger={\frac {1}{\sqrt{2m\hbar\omega}}}(-ip+m\omega x)
\end{equation}
are the corresponding annihilation and creation operators, respectively. It is
noticed that in Ref.~\cite{1}, $m$,$\omega$ and $\hbar$ all are unit.  In
order to consider thermal effects, thermofield dynamics introduces a copy of
the physical oscillator Eq.(1) ( called the tilde oscillator )
\begin{equation}
\tilde{H}={\frac {1}{2m}}\tilde{p}^2 +{\frac {1}{2}} m\omega^2 \tilde{x}^2
  =(\tilde{a}^\dagger \tilde{a}+{\frac {1}{2}})\hbar\omega \;,
\end{equation}
according to the tilde ``conjugation'': $\widetilde{C O}\equiv C^* \tilde{O}$
\cite{16}. Here, $C$ is any coefficient appeared in expressions of quantities
for the physical system, $O$ any operator, the superscript $*$ means complex
conjugation, and $\tilde{O}$ represents the corresponding operator for the
tilde system. Exploiting the physical and tilde oscillators, one can have
the thermal vacuum \cite{16}
\begin{equation}
|0,\beta>=T(\theta)|0,\tilde{0}>  \;,
\end{equation}
where, $|0,\tilde{0}>=|0>|\tilde{0}>$ is the product of ground states of the
physical and tilde oscillators, $\beta={\frac {1}{k_b T}}$ with $k_b$ the
Boltzmann constant and $T$ the temperature, and the unitary transformation
$T(\theta)$ ( called thermal transformation ) is
\begin{equation}
T(\theta)=exp\{-\theta(\beta)(a\tilde{a}-a^\dagger \tilde{a}^\dagger)\}
\end{equation}
with $$tanh[\theta(\beta)]=e^{-\beta\hbar\omega/2} .$$ Notice that any
physical operator commutes with any tilde operator. Consequently the
thermal-vacuum average value agrees with canonical ensemble average in
statistical mechanics.

It is evident that the thermal vacuum (5) is similar to the two-mode squeezed
states discussed in Ref.~\cite{8} except for a minus difference between the
exponents in Eq.(6) here and Eq.(37) there. Although the wavefunction of the
two-mode squeezed state was given in the coordinate representation \cite{8},
here we still derive the position wavefunction of the thermal vacuum for the
sake of both the completeness and the establishment of the thermal coordinate
representation. Substituting Eq.(3) into Eq.(6), one can read
\begin{equation}
T(\theta)=exp\{i{\frac {\theta}{\hbar}}(x\tilde{p}- \tilde{x} p)\}
\end{equation}
with $\theta\equiv \theta(\beta)$. From Appendix B.4 in Ref.~\cite{26}, the
last formula can be unentangled as
\begin{equation}
T(\theta)=exp\{-tanh(\theta)\tilde{x}\partial_x\}
          exp\{ln[cosh(\theta)](x\partial_x -\tilde{x}\partial_{\tilde{x}})\}
          exp\{-tanh(\theta)x\partial_{\tilde{x}}\} \;.
\end{equation}
Using the following operator properties \cite{27}
\begin{equation}
e^{C\partial_y }f(y)=f(y+C)
\end{equation}
and
\begin{equation}
e^{C y\partial_y}f(y)=f(ye^C) ,
\end{equation}
which are proved easily, we obtain the wavefunction of the thermal vacuum as
\begin{eqnarray}
<\tilde{x},x|0,\beta> &=& T(\theta)({\frac {m\omega}
           {\pi \hbar}})^{\frac {1}{2}} exp\{-{\frac {m\omega}{2\hbar}}
           (x^2+\tilde{x}^2)\} \nonumber  \\
      &=& ({\frac {m\omega}{\pi \hbar}})^{\frac {1}{2}}
    exp\{-{\frac {m\omega}{2\hbar}}[(x cosh(\theta)-\tilde{x} sinh(\theta))^2
        +(\tilde{x} cosh(\theta)-x sinh(\theta))^2]\} \;.
\end{eqnarray}
When $\beta\to\infty$, $<\tilde{x},x|0,\beta>$ is reduced to $<\tilde{x},x
|0,\tilde{0}>$. This expression Eq.(11) can be generalized to the Gaussian
wavefunctional approach for equilibrium field theory in thermofield dynamics
\cite{28}. 

 Such an expression of the thermal vacuum wavefunction Eq.(11) suggests the
usefulness of introducing a thermal coordinate representation. In thermofield
dynamics, for any operator of the physical or tilde oscillator $Q$, its
thermal counterpart is defined as $Q_\beta \equiv T(\theta)Q T^\dagger
(\theta)$ \cite{16}. In particular, for the fundamental canonical conjugate
pairs $\{x,p=-i\hbar\partial_x\}$ and $\{\tilde{x},\tilde{p}
=i\hbar\partial_{\tilde{x}}\}$, the corresponding thermal operators are
\begin{equation}
x_\beta\equiv T(\theta)xT^\dagger (\theta)=x cosh(\theta)-
        \tilde{x}sinh(\theta)
,  \;\;\; p_\beta\equiv T(\theta)pT^\dagger (\theta)=p cosh(\theta)
         -\tilde{p}sinh(\theta)
\end{equation}
and
\begin{equation}
\tilde{x}_\beta\equiv T(\theta)\tilde{x}T^\dagger (\theta)
  =\tilde{x} cosh(\theta)-x sinh(\theta),  \;\;\;
\tilde{p}_\beta\equiv T(\theta)pT^\dagger (\theta)
  =\tilde{p} cosh(\theta)-p sinh(\theta) \;.
\end{equation}
Obviously, the thermal vacuum wavefunction Eq.(11) can be written as
\begin{equation}
<\tilde{x},x|0,\beta> = ({\frac {m\omega}{\pi \hbar}})^{\frac {1}{2}}
      exp\{-{\frac {m\omega}{2\hbar}}(x_\beta^2 + \tilde{x}_\beta^2)\} \;,
\end{equation}
which is the same form with the wavefunction $<\tilde{x},x|0,0>$. Noticing
that the commutators $[x_\beta, p_\beta]=i\hbar$, $[\tilde{x}_\beta,
\tilde{p}_\beta]=-i\hbar$ and $[O_\beta, \tilde{O}_\beta]=0$ hold, one can set
$p_\beta\equiv -i\hbar{\frac {\partial}{\partial x_\beta}}$ and
$\tilde{p}_\beta\equiv i\hbar{\frac {\partial}{\partial \tilde{x}_\beta}}$,
and establish a representation for the thermal oscillator, in which any
object ( operators, wavefunctions ) can be expressed in terms of $x_\beta$,
$\tilde{x}_\beta$, ${\frac {\partial}{\partial x_\beta}}$ and/or
${\frac {\partial}{\partial \tilde{x}_\beta}}$. In this paper, we shall call
it thermal coordinate representation. Evidently, this representation is
reached through the unitary thermal transformation $T(\theta)$ of the
coordinate representation. When working in the representation,
quantities will take similar forms to those in quantum mechanics, and hence it
will simplify our derivation in the present paper.

It is suitable here to mention a mathematical property and the physical sense
of the thermal transformation Eq.(6) . It is shown easily that the action of
$T(\theta)$ on a function of the physical and tilde positions
$\{x,\tilde{x}\}$ amounts to just the thermal coordinate $x_\beta,
\tilde{x}_\beta$ taking the place of $x, \tilde{x}$, that is,
\begin{equation}
T(\theta)f(x,\tilde{x})=f(x_\beta,\tilde{x}_\beta)  \;.
\end{equation}
Thermal transformation is also called thermalizing operator \cite{18}. It
describes the effect of a thermal reservoir in which a quantum harmonic
oscillator immerses. From Eq.(5), we can say loosely that a thermalizing
operator heats the ground state of a zero-temperature harmonic oscillator into
a thermal vacuum with a finite temperature. In quantum optics, the
thermalizing operator describes the action of a source which excites one-mode
electromagnetic field from its ground state to a chaotic state ( thermalized
radiation ). Thus, in order to consider thermal noise, it is enough to perform
the action of the thermalizing operator on the non-classical states mentioned
in the last section. Next, we shall address them.

\section{Thermalized Displaced Number and Squeezed Number State
              in the Coordinate Representation}
\label{3}

\addtocounter{equation}{15}

  Because both coherent and squeezed states are constructed with the
displacing operator and squeezing operator acting on the ground state, there
are three different states with squeezed effect : squeezed state ( only the
squeezing operator acting on the ground state ), displaced squeezed state, and
squeezed displaced state, which are all usually called squeezed state in the
literature. In this paper, the terminology ``squeezed state'' means only the
displaced squeezed state, for which the action of the displacing operator
follows that of the squeezing operator. So does the squeezed number state.
However, when introducing a finite temperature effect, one still faces more
choices about the orders among displacing, squeezing and thermalizing. A
different order will lead to a different thermal non-classical state \cite{18}.
Nevertheless, if using thermal creation and annihilation operators to work,
$i.e.$, doing as done in Ref.~\cite{17,19}, one can escape the order problems
with thermalizing operator. In this section, we shall introduce a finite
temperature effect into the displaced number state and squeezed number state
by using the thermal creation and annihilation operators with the vacuum
$|0,\beta>$ \cite{17,19} and then give their expressions in the coordinate
representation. This construction is utterly to thermalize the displaced
number and squeezed number states, namely, it gives the thermalized
displaced number and squeezed number states, as one shall be seen later.

The thermal annihilation and creation operators with the thermal vacuum Eq.(5)
are \cite{16}
\begin{equation}
a_\beta=T(\theta)aT^\dagger(\theta), a^\dagger_\beta=T(\theta)a^\dagger
        T^\dagger(\theta)
\end{equation}
and
\begin{equation}
\tilde{a}_\beta=T(\theta)\tilde{a}T^\dagger(\theta), \tilde{a}^\dagger_\beta=
        T(\theta)\tilde{a}^\dagger T^\dagger(\theta) \;.
\end{equation}
One can easily check that $a_\beta|0,\beta>=0$, $\tilde{a}_\beta|0,\beta>=0$
and $[a_\beta,a^\dagger_\beta]=[\tilde{a}_\beta,\tilde{a}^\dagger_\beta]=1$.
With the aids of thermal creation operators $a^\dagger_\beta$ and
$\tilde{a}^\dagger_\beta$, one can construct normalized thermal number states
\begin{equation}
|n,m,\beta>=
{\frac {1}{\sqrt{n!m!}}} a^{\dagger \; n}_\beta \tilde{a}^{\dagger \; m}_\beta
            |0,\beta>
\end{equation}
with the closure relation
\begin{equation}
\sum_{n,m}|n,m,\beta><\beta,m,n|=1 \;.
\end{equation}

The so-called displaced number state $|\alpha,n>$ of the oscillator Eq.(1)
is defined in Fock space as $|\alpha,n> \equiv D(\alpha)|n>$ and can
have the following form in the coordinate representation \cite{1}
\begin{eqnarray}
<x|\alpha,n> &=& ({\frac {m\omega}{\pi \hbar}})^{\frac {1}{4}}
                 {\frac {1}{\sqrt{2^n n!}}} exp\{-i\alpha_1 \alpha_2\}
                 \nonumber \\  &\;\;\;&
              \cdot exp\{-{\frac {m\omega}{2\hbar}}(x -
               \sqrt{{\frac {2\hbar}{m\omega}}}\alpha_1)^2
               + i \sqrt{{\frac {2 m\omega}{\hbar}}} \alpha_2 x \}
            H_n[\sqrt{{\frac {m\omega}{\hbar}}}x-\sqrt{2}\alpha_1]
\end{eqnarray}
with $\alpha=(\alpha_1+i\alpha_2)$ any complex number, $H_n[\cdots]$ the
Hermite polynomials and the displacing operator
\begin{equation}
D (\alpha)=e^{\alpha a^\dagger-\alpha^* a} \;.
\end{equation}
In this definition, when the number state $|n>$ is replaced by the ground
state $|0>$, the state $|\alpha,n>$ is reduced to the usual coherent state
$|\alpha>$. Evidently, the state $|\alpha,n>$ is constructed just with the
displacing operator $D(\alpha)$ acting on the number state $|n>$. Similarly,
one can define the following state $|\alpha,n,\beta>$
\begin{equation}
|\alpha,n,\beta>=D_\beta(\alpha)\tilde{D}_\beta(\alpha)|n,n,\beta>
\end{equation}
so as to introduce a finite temperature effect into the displaced number state.
Here, the thermal displacing operators $D_\beta (\alpha)$ and
$\tilde{D}_\beta (\alpha)$ are
\begin{equation}
D_\beta (\alpha)=exp\{\alpha a_\beta ^\dagger-\alpha^* a_\beta \}
\end{equation}
\begin{equation}
\tilde{D}_\beta (\alpha)=exp\{\tilde{\alpha} \tilde{a}^\dagger_\beta
                         -\tilde{\alpha}^* \tilde{a}_\beta\} \;,
\end{equation}
respectively, which are generalization of the displacing operator $D(\alpha)$.
Note that $\tilde{\alpha}=\alpha^*$ in the present paper ( Of course, one can
take $\tilde{\alpha}$ as another parameter independent of $\alpha$ ). When
$n=0$ the state $|\alpha,n,\beta>$ is just Eq.(11) with $\gamma=\alpha$ in
Ref.~\cite{19}( the first paper ) and Eq.(3.1) with $\varphi=\alpha$ in
Ref.~\cite{17}. Employing the definitions (5), (16), (17), (18), we obtain
\begin{equation}
|\alpha,n,\beta> = T(\theta)|\alpha,n>|\tilde{\alpha},n> \;.
\end{equation}
This equation indicates that the state $|\alpha,n,\beta>$ is just the
thermalized displaced number state. In the last equation, $|\tilde{\alpha},n>$
is the tilde displaced number state and can be obtained from Eq.(20) according
to the tilde rules. When $n=0$, the state $|\alpha,0,\beta>$ is the
thermalized coherent state, being similar to Eq.(3.3) in Ref.~\cite{17}.
Employing Eqs.(15),(12) and (13), one can have
\begin{eqnarray}
<\tilde{x},x|\alpha,n,\beta> &=& ({\frac {m\omega}{\pi \hbar}})^{\frac {1}{2}}
                 {\frac {1}{2^n n!}}exp\{-{\frac {m\omega}{2\hbar}}
                 [(x_\beta - \sqrt{{\frac {2\hbar}{m\omega}}}\alpha_1)^2
             + (\tilde{x}_\beta - \sqrt{{\frac {2\hbar}{m\omega}}}\alpha_1)^2]
           \nonumber \\ &\;\;\;&
 + i \sqrt{{\frac {2 m\omega}{\hbar}}} \alpha_2 (x_\beta-\tilde{x}_\beta) \}
         H_n[\sqrt{{\frac {m\omega}{\hbar}}}x_\beta-\sqrt{2}\alpha_1]
      H_n[\sqrt{{\frac {m\omega}{\hbar}}}\tilde{x}_\beta-\sqrt{2}\alpha_1] \\
           &=& ({\frac {m\omega}{\pi \hbar}})^{\frac {1}{2}}
                 {\frac {1}{2^n n!}}exp\{-{\frac {m\omega}{2\hbar}}
                 [(x cosh(\theta)-\tilde{x}sinh(\theta)
                 - \sqrt{{\frac {2\hbar}{m\omega}}}\alpha_1)^2  \nonumber \\
   &\;\;\;&          + (\tilde{x}cosh(\theta)-x sinh(\theta)
              - \sqrt{{\frac {2\hbar}{m\omega}}}\alpha_1)^2]
  + i \sqrt{{\frac {2 m\omega}{\hbar}}} \alpha_2 (cosh(\theta)+sinh(\theta))
             (x-\tilde{x}) \}  \nonumber \\  &\;\;\;&   \cdot
  H_n[\sqrt{{\frac {m\omega}{\hbar}}}(x cosh(\theta)-\tilde{x}sinh(\theta))
           -\sqrt{2}\alpha_1]
 H_n[\sqrt{{\frac {m\omega}{\hbar}}}(\tilde{x}cosh(\theta)-x sinh(\theta))
      -\sqrt{2}\alpha_1] \; .
\end{eqnarray}
The $r.h.s$ of Eq.(26) is the wavefunction of the thermalized displaced number
state in the thermal coordinate representation, and Eq.(27) is just the
wavefunction in the coordinate representation. When $\beta\to\infty$, Eq.(27)
is reduced to a product of $x$-function and $\tilde{x}$-function, each factor
resembling Eq.(15) in Ref.~\cite{1}.

Now we are at the position to discuss the thermalized squeezed number state.
The squeezed number state $|\alpha,z,n>$ of the oscillator Eq.(1) is
constructed by using the squeezing operator $S(z)$ \cite{4}
\begin{equation}
S(z)=exp\{-{\frac {1}{2}}(z^* a a - z a^\dagger a^\dagger)\}
\end{equation}
and reads
\begin{equation}
|\alpha,z,n> \equiv D(\alpha)S(z)|n>  \;.
\end{equation}
Here, $z$ is any complex constant. When $n=0$, $|\alpha,z,0>$ is the usual
squeezed state. From Ref.~\cite{1}, the wavefunction of $|\alpha,z,n>$ in the
coordinate representation is ( here in terms of our notations )
\begin{eqnarray}
<x|\alpha,z,n> &=& ({\frac {m\omega}{\pi \hbar}})^{\frac {1}{4}}
             {\frac {(\sqrt{{\cal F}_3})^{n}}{\sqrt{{\cal F}_1 2^n n!}}}
        exp\{-i\alpha_1\alpha_2\} \nonumber  \\
    &\;\;\;&  \cdot  exp\{-{\frac {m\omega}{2 \hbar}}{\cal F}_2
         [x-\sqrt{{\frac {2\hbar}{m\omega}}}\alpha_1]^2
         +i\sqrt{{\frac {2m\omega}{\hbar}}}\alpha_2 x\}
         H_n[\sqrt{{\frac {m\omega}{\hbar}}}({\cal F}_4)^{-1}
         (x-\sqrt{{\frac {2\hbar}{m\omega}}}\alpha_1)]
\end{eqnarray}
with $z=z_1+iz_2=r e^{i\phi}$, ${\cal S}=cosh(r)+z_1 sinh(r)/r$ and
$\kappa=z_2 sinh(r)/(2r{\cal S})$. In Eq.(30), for the convenience of
comparison later, we adopted the notations ${\cal F}$'s in Ref.~\cite{1},
that is,
\begin{eqnarray}
{\cal F}_1&=&{\cal S}(1+i2\kappa), \;\;
{\cal F}_2={\frac {1}{{\cal S}^2(1+i2\kappa)}}-i2\kappa, \nonumber \\
{\cal F}_3&=&{\frac {1-i2\kappa}{1+i2\kappa}}, \ \ \ \ \
{\cal F}_4={\cal S}(1+4\kappa^2)^{\frac {1}{2}} \; .
\end{eqnarray}
In analogy with the definition of the squeezed number state, we introduce the
thermal squeezing operator
\begin{equation}
S_\beta(z)=exp\{-{\frac {1}{2}}(z^* a_\beta a_\beta
         - z a_\beta^\dagger a_\beta^\dagger)\}
           \;, \;\;\;\;\;
\tilde{S}_\beta(z)
          =exp\{-{\frac {1}{2}}(\tilde{z}^* \tilde{a}_\beta \tilde{a}_\beta
         - z \tilde{a}_\beta^\dagger  \tilde{a}_\beta^\dagger)\}
\end{equation}
and define the following state $|\alpha,z,n,\beta>$
\begin{equation}
|\alpha,z,n,\beta> \equiv D_\beta(\alpha)
       \tilde{D}_\beta(\alpha)S_\beta(z)\tilde{S}_\beta(z)|n,n,\beta> \;.
\end{equation}
to introduce a finite temperature effect. Note that $\tilde{z}=z^*$ in the
present paper ( Of course, one can take $\tilde{z}$ as another parameter
independent of $z$ ). It is easily shown that
\begin{equation}
|\alpha,z,n,\beta>=T(\theta)|\alpha,z,n>|\tilde{\alpha},\tilde{z},n>
\end{equation}
with $|\tilde{\alpha},\tilde{z},n>$ the tilde version of $|\alpha,z,n>$.
Evidently, the last equation indicates that the state $|\alpha,z,n,\beta>$ is
just the thermalized displaced number state. When $n=0$, $|\alpha,z,0,\beta>$
is just the thermalizd squeezed state, being similar to Eq.(11) in
Ref.~\cite{19}( the second paper ). Employing Eqs.(15),(30),(12) and (13), we
obtain
\begin{eqnarray}
<\tilde{x},x|\alpha,z,n,\beta>&=& ({\frac {m\omega}{\pi \hbar}})^{\frac {1}{2}}
    {\frac {|{\cal F}_3|^{n}}{|{\cal F}_1| 2^n n!}}
    exp\{-{\frac {m\omega}{2 \hbar}}
    [{\cal F}_2(x_\beta-\sqrt{{\frac {2\hbar}{m\omega}}}\alpha_1)^2
    \nonumber  \\  &\;\;\;&   +
   {\cal F}_2^*(\tilde{x}_\beta-\sqrt{{\frac {2\hbar}{m\omega}}}\alpha_1)^2 ]
      +i\sqrt{{\frac {2m\omega}{\hbar}}}\alpha_2 (x_\beta-\tilde{x}_\beta)\}
         \nonumber \\ & \ \ \ &
      \cdot H_n[\sqrt{{\frac {m\omega}{\hbar}}}({\cal F}_4)^{-1}
         (x_\beta-\sqrt{{\frac {2\hbar}{m\omega}}}\alpha_1)]
         H_n[\sqrt{{\frac {m\omega}{\hbar}}}({\cal F}_4)^{-1}
         (\tilde{x}_\beta-\sqrt{{\frac {2\hbar}{m\omega}}}\alpha_1)] \\
    &=& ({\frac {m\omega}{\pi \hbar}})^{\frac {1}{2}}
    {\frac {|{\cal F}_3|^{n}}{|{\cal F}_1| 2^n n!}}
    exp\{-{\frac {m\omega}{2 \hbar}}
    [{\cal F}_2(x cosh(\theta)-\tilde{x} sinh(\theta)-
    \sqrt{{\frac {2\hbar}{m\omega}}}\alpha_1)^2   \nonumber  \\  & \ \ \ & +
   {\cal F}_2^*(\tilde{x}cosh(\theta)-x sinh(\theta)-
   \sqrt{{\frac {2\hbar}{m\omega}}}\alpha_1)^2 ] \nonumber  \\  & \ \ \ &
      +i\sqrt{{\frac {2m\omega}{\hbar}}}\alpha_2 (cosh(\theta)+sinh(\theta))
      (x-\tilde{x})\}
         \nonumber \\ & \ \ \ &
      \cdot H_n[\sqrt{{\frac {m\omega}{\hbar}}}({\cal F}_4)^{-1}
            (x cosh(\theta)-\tilde{x} sinh(\theta)-
            \sqrt{{\frac {2\hbar}{m\omega}}}\alpha_1)]
            \nonumber \\ &\;\;\;&
      \cdot H_n[\sqrt{{\frac {m\omega}{\hbar}}}({\cal F}_4)^{-1}
         (\tilde{x}cosh(\theta)-x sinh(\theta)-
         \sqrt{{\frac {2\hbar}{m\omega}}}\alpha_1)] \;.
\end{eqnarray}
The expression Eq.(36) is just the wavefunction of the thermalized squeezed
number state in the coordinate representation. When $\beta\to\infty$, the
x-part of Eq.(36) is consistent with Eq.(20) in Ref.~\cite{1}.

In this section, we have constructed the thermalized displaced number and
squeezed number states with the thermal creation and annihilation operators,
and given their wavefunctions. These states are physically meaningful. For a
displaced thermalized squeezd state, Fearn and Collett gave a physical
interpretation that it corresponds to the output from a linear photon
amplifier whose input is a squeezed state if the amplifier's added noise is
regarded as thermal photons \cite{18}. Thus, according to this interpretation,
it is not difficult to give physical interpretations for the states in the
present paper : thermalized displaced number and squeezed number states. For
instance, the thermalized displaced number state should correspond, at least
theoretically, to the output from a thermal source who excites the previous
output from a linear photon amplifier with a number state being the input, if
no thermal noises company the amplifier and the thermal source can excite an
electromagnetic field from its ground state to a thermal chaotic state. Of
course, strictly speaking, it is impossible to have no thermal noises, and so
one should consider a type of completely thermalizied state in which both the
displacing and the squeezing are companied by thermal noises. We shall discuss
the more practical situations in a separate paper. Next, we shall discuss the
time evolution of the two thermalized non-classical states.

\section{Time-evolution of the Thermalized Non-classical States}
\label{4}

\addtocounter{equation}{36}

In this section, we first consider the time-evolution of the thermalized
displaced number and squeezed number states and then calculate position
probability densities of them.

In thermofield dynamics, Hamiltonian $\hat{H}$ of the combined system of the
physical and tilde oscillators is \cite{16}
\begin{equation}
\hat{H}=H-\tilde{H}=(a^\dagger
a-\tilde{a}^\dagger \tilde{a})\hbar\omega=(a_\beta^\dagger
     a_\beta-\tilde{a}_\beta^\dagger \tilde{a}_\beta) \hbar\omega    \;.
\end{equation}
Hence the time-evolution operator of the combined system is \cite{16}
\begin{equation}
U(t)=exp\{-{\frac {i}{\hbar}}\hat{H}t\}=exp\{-i\omega t a_\beta^\dagger
     a_\beta\}exp\{i\omega t \tilde{a}_\beta^\dagger \tilde{a}_\beta)\}
\end{equation}
with $t$ the time. From Eqs.(3), (16) and (17), we have
\begin{equation}
a_\beta={\frac {1}{\sqrt{2m\hbar\omega}}}(ip_\beta+m\omega x_\beta)
       \;, \;\;\;\;\;
a_\beta^\dagger={\frac {1}{\sqrt{2m\hbar\omega}}}(-ip_\beta+m\omega x_\beta)
\end{equation}
and
\begin{equation}
\tilde{a}_\beta={\frac {1}{\sqrt{2m\hbar\omega}}}(-i\tilde{p}_\beta
                +m\omega\tilde{x}_\beta) \;, \;\;\;\;\;
\tilde{a}_\beta^\dagger={\frac {1}{\sqrt{2m\hbar\omega}}}(i\tilde{p}_\beta
                +m\omega \tilde{x}_\beta) \;.
\end{equation}
Thus, in the thermal coordinate representation, the time-evolution operator
can be unentangled as
\begin{eqnarray}
U(t)&=&{\frac {1}{cos(\omega t)}}
     exp\{-i{\frac {m\omega}{2\hbar}}tan(\omega t)x_\beta^2\}
     exp\{-log(cos(\omega t))x_\beta {\frac {\partial}{\partial x_\beta}}\}
     exp\{i{\frac {\hbar}{2m\omega}}tan(\omega t)
           {\frac {\partial^2}{\partial x_\beta^2}}\}
         \nonumber \\  & \ \ \ & \cdot
     exp\{i{\frac {m\omega}{2\hbar}}tan(\omega t)\tilde{x}_\beta^2\}
     exp\{-log(cos(\omega t))\tilde{x}_\beta
          {\frac {\partial}{\partial \tilde{x}_\beta}}\}
     exp\{-i{\frac {\hbar}{2m\omega}}tan(\omega t)
           {\frac {\partial^2}{\partial \tilde{x}_\beta^2}}\} .
\end{eqnarray}
The operator $U(t)$ acting on a wavefunction will yield the time-evolution of
the wavefunction. With the help of Eqs.(9) and (10) and the following operator
property \cite{27}
\begin{equation}
exp\{C \partial^2_y\}f(y)={\frac {1}{\sqrt{4\pi C}}}
        \int^{\infty}_{-\infty}exp\{-{\frac {(w-y)^2}{4C}}\}f(w)dw \;,
\end{equation}
which can be shown by using the identity
$$ e^{C Q^2}={\frac {1}{\sqrt{2\pi}}} \int^{\infty}_{-\infty} e^{-\sigma^2/2}
              e^{\sigma \sqrt{2 C} Q} d\sigma , $$
we can perform the action of $U(t)$ on various thermalized non-classical
states here.

Leting $U(t)$ Eq.(41) act on the wavefunction Eq.(14), one can find that
the wavefunction Eq.(14) is invariant and hence the wavefunction of the
thermal vacuum is independent of time, $i.e.,$ $<\tilde{x},x|0,\beta,t>\equiv
U(t)<\tilde{x},x|0,\beta>=<\tilde{x},x|0,\beta>$. This is understandable
because the average value of any physical observable on the thermal vacuum is
equal to its ensemble average value, which does not vary with time.

$U(t)$ Eq.(41) acting on Eq.(26) gives the time-dependent wavefunction of the
thermalized displaced number state $<\tilde{x},x|\alpha,n,\beta,t>\equiv
U(t)<\tilde{x},x|\alpha,n,\beta>$ as
\begin{eqnarray}
<\tilde{x},x|\alpha,n,\beta,t>&=&({\frac {m\omega}{\pi \hbar}})^{\frac {1}{2}}
                 {\frac {1}{2^n n!}} exp\{({\frac {\alpha^2}{A}}
                 +{\frac {{\alpha^*}^2}{A^*}})cos(\omega t) -2\alpha_1^2\}
                 \nonumber  \\  & \ \ \ &
            \cdot exp\{-{\frac {m\omega}{2\hbar}}
      [(x_\beta - \sqrt{{\frac {2\hbar}{m\omega}}}{\frac {\alpha}{A}})^2 +
(\tilde{x}_\beta - \sqrt{{\frac {2\hbar}{m\omega}}}{\frac {\alpha^*}{A^*}})^2]
         \}  \nonumber \\ &\;\;\;&
     \cdot H_n[\sqrt{{\frac {m\omega}{\hbar}}}x_\beta
        -\sqrt{2}(\alpha_1 cos(\omega t)+\alpha_2 sin(\omega t))]
      H_n[\sqrt{{\frac {m\omega}{\hbar}}}\tilde{x}_\beta
      -\sqrt{2}(\alpha_1 cos(\omega t)+\alpha_2 sin(\omega t))]
\end{eqnarray}
with $A=cos(\omega t)+i sin(\omega t)$. In finishing the relevant integral
with Hermite polynomial, we used the formula 7.374 (8) in Ref.~\cite{29}.
Eq.(43) indicates that the wavefunction of the thermalized displaced number
state is dependent upon time.

Similarly, $U(t)$ Eq.(41) acting on Eq.(35) yields the time-dependent
wavefunction of the thermalized squeezed number state
$<\tilde{x},x|\alpha,z,n,\beta,t>\equiv U(t)<\tilde{x},x|\alpha,z,n,\beta>$ as
\begin{eqnarray}
<\tilde{x},x|\alpha,z,n,\beta,t>&=&
    ({\frac {m\omega}{\pi \hbar}})^{\frac {1}{2}}
    {\frac {|{\cal F}_3|^{n}}{|{\cal F}_1|2^n n!|B|}}
    \nonumber \\ &\;\;\;& \cdot
    exp\{-{\frac {{\cal F}_2 cos(\omega t) \alpha_1^2
  + 2 {\cal F}_2 sin(\omega)\alpha_1 \alpha_2+i sin(\omega t)\alpha_2^2}{B}}\}
    \nonumber \\ &\;\;\;&   \cdot
    exp\{-{\frac {{\cal F}_2^* cos(\omega t) \alpha_1^2
    + 2 {\cal F}_2^* sin(\omega)
    \alpha_1 \alpha_2-i sin(\omega t)\alpha_2^2}{B^*}}\}
    \nonumber \\ &\;\;\;&  \cdot
    exp\{-{\frac {m\omega}{2 \hbar}}
    {\frac {{\cal F}_2 cos(\omega t)+i sin(\omega t)}{B}}x_\beta^2+
 2 \sqrt{{\frac {m\omega}{2\hbar}}}{\frac {{\cal F}_2 \alpha_1+i\alpha_2}{B}}
 x_\beta \}
    \nonumber \\ &\;\;\;& \cdot
    exp\{-{\frac {m\omega}{2 \hbar}}
  {\frac {{\cal F}_2^* cos(\omega t)-i sin(\omega t)}{B^*}}\tilde{x}_\beta^2+
        2 \sqrt{{\frac {m\omega}{2\hbar}}}
      {\frac {{\cal F}_2^* \alpha_1-i\alpha_2}{B^*}}\tilde{x}_\beta \}
    \nonumber  \\  &\;\;\;&    \cdot
         H_n[\sqrt{{\frac {m\omega}{\hbar}}}({\cal F}_4|B|)^{-1}
         (x_\beta-\sqrt{{\frac {2\hbar}{m\omega}}}
         (cos(\omega t)\alpha_1+sin(\omega t)\alpha_2))]
         \nonumber \\ &\;\;\;& \cdot
         H_n[\sqrt{{\frac {m\omega}{\hbar}}}({\cal F}_4|B|)^{-1}
         (\tilde{x}_\beta-\sqrt{{\frac {2\hbar}{m\omega}}}
         (cos(\omega t)\alpha_1+sin(\omega t)\alpha_2))]  \;.
\end{eqnarray}
with $B=cos(\omega t)+i {\cal F}_2 sin(\omega t)$. Like the thermalized
displaced number state, the wavefunction of the thermalized squeezed number
state is also time-dependent. Substituting Eqs.(12) and (13) into Eqs.(43) and
(44), one can obtain the time-dependent wavefunctions of the thermalized
displaced number and squeezed number states in the coordinate representation.
When $\beta\to \infty$, the x-part of Eq.(44) is consistent with Eq.(45) in
Ref.~\cite{1}, except for the lack of an imaginary exponent, which is
cancelled by the relevant exponent from the tilde system ( In Eq.(46) of
Ref.~\cite{1} there should be the symbol ``='' between the parentheses and the
fraction ). Additionally, for the case of $\beta\to\infty$, the x-part of
Eq.(43) is consistent with the zero-$z$ resultant of Eq.(45) in Ref.~\cite{1}
( except for the lack of an imaginary exponent ).

Eqs.(43) and (44) indicate that the wavefunctions of both the thermalized
displaced number state and the thermalized squeezed number state are dependent
upon time, which is different from the thermal vacuum wavefunction. This point
is because both the thermalized displaced number state and the thermalized
squeezed number state are not eigenstates of the Hamiltonian $\hat{H}$, while
the thermal vacuum is an eigenstate of the Hamiltonian $\hat{H}$ with zero
eigenvalue.

Now we can calculate the position probability densities. The probability
density is the modulus square of the position-coordinate wavefunction with the
tilde coordinate integrated. At the first, we consider the thermal vacuum.
The density is easy to calcuate and the result is
\begin{eqnarray}
\rho_v(x,t)&\equiv& \int^\infty_{-\infty} <t,\beta,0|x,\tilde{x}>
                  <\tilde{x},x|0,\beta,t> d\tilde{x}
                =\int^\infty_{-\infty}<\beta,0|x,\tilde{x}>
                   <\tilde{x},x|0,\beta> d\tilde{x} \nonumber  \\
   &=& \sqrt{{\frac {m\omega}{\pi \hbar}}} tanh^{{\frac {1}{2}}}
   ({\frac {\beta\hbar\omega}{2}}) exp\{- {\frac {m\omega}{\hbar}}
   tanh({\frac {\beta\hbar\omega}{2}}) x^2\} \;.
\end{eqnarray}
This is just the familiar result in statistical mechanics.

Secondly, we calculate the position probability density of the thermalized
displaced number state
\begin{equation}
\rho_c(x,t)\equiv \int^{\infty}_{-\infty}
   <t,\beta,n,\alpha|x,\tilde{x}><\tilde{x},x|\alpha,n,\beta,t> d\tilde{x}\;.
\end{equation}
Substituting Eqs.(12),(13) and (43) into Eq.(46) and reducing it, we have
\begin{equation}
\rho_c(x,t)={\frac {m\omega}{\pi \hbar}}({\frac {1}{2^n n!}})^2
    \int^{\infty}_{-\infty} exp\{-(a_1\tilde{x}+a_2)^2-(b_1\tilde{x}+b_2)^2\}
    (H_n[a_1\tilde{x}+a_2])^2(H_n[-b_1\tilde{x}-b_2])^2 d\tilde{x}
\end{equation}
where,
\begin{eqnarray*}
a_1=\sqrt{{\frac {m\omega}{\hbar}}} cosh(\theta),\;\;\;\;\;
a_2=-\sqrt{{\frac {m\omega}{\hbar}}} x sinh(\theta)
    -\sqrt{2}(\alpha_1 cos(\omega t)+\alpha_2 sin(\omega t))  \\
b_1=\sqrt{{\frac {m\omega}{\hbar}}} sinh(\theta),\;\;\;\;\;
b_2=-\sqrt{{\frac {m\omega}{\hbar}}} x cosh(\theta)
    +\sqrt{2}(\alpha_1 cos(\omega t)+\alpha_2 sin(\omega t))  \;.
\end{eqnarray*}
With the help of the formula on page 225 of Ref.~\cite{30}
\begin{equation}
H_m[x]H_n[x]=\sum^{min\{m,n\}}_{r=0} 2^r r! C_m^r C^r_n H_{m+n-2r}[x]
\end{equation}
with $C_n^r$ and $C_m^r$ being combinations, Eq.(47) can be written as
\begin{equation}
\rho_c(x,t)={\frac {m\omega}{\pi \hbar a_1}}({\frac {1}{2^n n!}})^2
    \sum^n_{j,k=0} 2^{j+k} j! k! (C_n^j C_n^k)^2 \int^{\infty}_{-\infty}
    exp\{-y^2-(a y+b)^2\} H_{2n-2j}[y]H_{2n-2k}[a y+b] dy
\end{equation}
with $a={\frac {b_1}{a_1}}$ and $b=-a a_2+b_2$.
Using repeatedly the formula
$$ e^{-x^2}H_n[x]=-{\frac {d}{d x}}\{e^{-x^2}H_{n-1}[x]\} $$
and then the formula 7.374(8) in Ref.~\cite{29}, one can obtain
\begin{eqnarray}
\rho_c(x,t) &=& {\frac {m\omega}{\pi \hbar a_1}}({\frac {1}{2^n n!}})^2
       \sqrt{{\frac {\pi}{a^2+1}}} exp\{-{\frac {b^2}{a^2+1}}\}
    \nonumber  \\  &\;\;\;& \cdot
    \sum^n_{j,k=0} 2^{j+k} j! k! (C_n^j C_n^k)^2 a^{2(n-j)}(a^2+1)^{j+k-2n}
     H_{2(2n-j-k)}[{\frac {b}{\sqrt{a^2+1}}}]   \\
&=& \sqrt{{\frac {m\omega}{\pi \hbar}}}({\frac {1}{2^n n!}})^2
    tanh^{{\frac {1}{2}}}({\frac {\beta\hbar\omega}{2}})
    \nonumber  \\  &\;\;\;&   \cdot
    exp\{-[\sqrt{{\frac {m\omega}{\hbar}}tanh({\frac {\beta\hbar\omega}{2}})}x
    -\sqrt{2}(\alpha_1 cos (\omega t)+\alpha_2 sin(\omega t))
    \sqrt{1+sech({\frac {\beta\hbar\omega}{2}})}]^2\} \nonumber \\ &\;\;\;&
    \cdot  \sum^n_{j,k=0} 2^{2(j+k)-2n} j! k! (C_n^j C_n^k)^2
     exp\{{\frac {1}{2}}(j-k)\hbar \omega \beta\}(cosh({\frac {\beta \hbar
     \omega}{2}}))^{j+k-2n} \nonumber \\ &\;\;\;&   \cdot
     H_{2(2n-j-k)}[\sqrt{{\frac {m\omega}{\hbar}}
     tanh({\frac {\beta\hbar\omega}{2}})}x
    -\sqrt{2}(\alpha_1 cos (\omega t)+\alpha_2 sin(\omega t))
    \sqrt{1+sech({\frac {\beta\hbar\omega}{2}})}]
\end{eqnarray}

As for the thermalized squeezed number state, the calculation of the position
probability density is completely similar to that of the thermalized displaced
number state. Finishing calculations similar to the above, one can have the
position probability density of the thermalized squeezed number state Eq.(44)
\begin{eqnarray}
\rho_s(x,t) & \equiv & \int^{\infty}_{-\infty}
         (|\alpha,z,n,\beta,t>)^*|\alpha,z,n,\beta,t> d\tilde{x}
          \\
&=& {\frac {m\omega {\cal F}_4|B|}{\pi \hbar a_1}}({\frac {1}{2^n n!}})^2
       \sqrt{{\frac {\pi}{a^2+1}}}
       {\frac {({\cal F}_3^*{\cal F}_3)^n}{{\cal F}_1^*{\cal F}_1B^*B}}
  exp\{-{\frac {b^2}{(a^2+1){\cal F}_4^2B^*B}}\} \nonumber \\ &\;\;\;& \cdot
    \sum^n_{j,k=0} 2^{j+k} j! k! (C_n^j C_n^k)^2 a^{2(n-j)}(a^2+1)^{j+k-2n}
     H_{2(2n-j-k)}[{\frac {b}{\sqrt{a^2+1}{\cal F}_4|B|}}] \\
&=& \sqrt{{\frac {m\omega}{\pi \hbar}}}({\frac {1}{2^n n!}})^2
   {\frac {1}{{\cal F}_4|B|}}
    tanh^{{\frac {1}{2}}}({\frac {\beta\hbar\omega}{2}})
   exp\{-{\frac {1}{{\cal F}_4^2B^*B}}
   \nonumber  \\  &\;\;\;&  \cdot   [
   \sqrt{{\frac {m\omega}{\hbar}}tanh({\frac {\beta\hbar\omega}{2}})}x
    -\sqrt{2}(\alpha_1 cos (\omega t)+\alpha_2 sin(\omega t))
    \sqrt{1+sech({\frac {\beta\hbar\omega}{2}})}]^2\} \nonumber \\ &\;\;\;&
   \cdot    \sum^n_{j,k=0} 2^{2(j+k)-2n} j! k! (C_n^j C_n^k)^2
     exp\{{\frac {1}{2}}(j-k)\hbar \omega \beta\}(cosh({\frac {\beta \hbar
     \omega}{2}}))^{j+k-2n} \nonumber \\ &\;\;\;&  \cdot
    H_{2(2n-j-k)}[{\frac {1}{{\cal F}_4|B|}}(
    \sqrt{{\frac {m\omega}{\hbar}}tanh({\frac {\beta\hbar\omega}{2}})}x
    -\sqrt{2}(\alpha_1 cos (\omega t)+\alpha_2 sin(\omega t))
    \sqrt{1+sech({\frac {\beta\hbar\omega}{2}})})] \;.
\end{eqnarray}
When $\beta\to\infty$, the existence of the factor $a^{2(n-j)}$ enforces the
summation index $j$ have a unique value $n$ because $a=0$. Thus, employing
Eq.(48), one has
$$\sum^n_{j,k=0} 2^{j+k} j! k! (C_n^j C_n^k)^2 a^{2(n-j)}(a^2+1)^{j+k-2n}
     H_{2(2n-j-k)}[{\frac {b}{\sqrt{a^2+1}{\cal F}_4|B|}}]
     = 2^n n!(H_n[{\frac {b}{\sqrt{a^2+1}{\cal F}_4|B|}}])^2  \;, $$
and hence Eq.(54) with $\beta\to\infty$ can give Eq.(50) in Ref.~\cite{1}.
Meanwhile, the probability density Eq.(51) with $\beta\to\infty$ can also lead
to Eq.(50) with $z=0$ in Ref.~\cite{1}.

For a given time, for example, $t=0$, one can get the position probability
densities $\rho_c(x)$ and $\rho_s(x)$ without considering the time evolution
from Eqs.(51) and (54). Making a contrast between $\rho_c(x)$ and $\rho_c(x,t)$
as well as $\rho_s(x)$ and $\rho_s(x,t)$, one can find that just the
displacement and squeeze parameters in the expressions of the densities
experience changes with the evolution of time. That is, for the thermalized
displaced number state, the real part of the displacement parameter $\alpha$
becomes $\alpha_1 cos(\omega t)+ i\alpha_2 sin(\omega t)$, which is similar to
that in Ref.~\cite{22}( 1965 ), and for the thermalized squeezed number state,
besides the same change of $\alpha_1$, the parameter ${\cal F}_4$ becomes
${\cal F}_4 |B|$.

The expressions of both $\rho_c(x,t)$ and $\rho_s(x,t)$ are complicated, but
one can easily prove that each of them is normalized, because only the term
with $j=k=n$ is not zero when $\rho_c(x,t)$ or $\rho_s(x,t)$ is integrated
with respect to $x$. Furthermore, one can calculate the average value of the
position coordinate $x$ on the thermalized squeezed number state as
\begin{equation}
<x>\equiv \int^{\infty}_{-\infty} x \rho_s(x,t)
           =\sqrt{coth({\frac {\beta\hbar\omega}{4}})}
   \sqrt{{\frac {2\hbar}{m\omega}}}(\alpha_1 cos (\omega t)+\alpha_2
             sin(\omega t)) \;.
\end{equation}
Evidently, $<x>$ is independent of $n$ and the squeeze parameter $z$, which is
similar to the result of the squeezed number state \cite{1}. Eq.(55) indicates
that the average value of $x$ on any thermalized squeezed number state follows
the motion of a classical harmonic oscillator, and the amplitude of the
oscillation increases with the increase of the temperature. When $\beta\to
\infty$, Eq.(55) is consistent with Eq.(36) in Ref.~\cite{1}, and contrasting
to Eq.(36) in Ref.~\cite{1}, Eq.(55) has just an additional temperature factor
$\sqrt{coth({\frac {\beta\hbar\omega}{4}})}$ .

Besides, one also easily obtain the variance of $x$ on a thermalized squeezed number state
as
\begin{equation}
(\Delta_n x)^2>\equiv \int^{\infty}_{-\infty} (x-<x>)^2 \rho_s(x,t)
     =coth({\frac {\beta\hbar\omega}{2}})(2n+1){\frac {\hbar{\cal F}_4^2 |B|^2
        }{2m\omega}} \;.
\end{equation}
When $\beta\to \infty$, $(\Delta_n x)^2$ is consistent with Eq.(38) in
Ref.~\cite{1}. When $n=0$, $(\Delta_0 x)^2$ is consistent with Eq.(15a) in
Ref.~\cite{19}( the second paper ). A comparison of Eq.(56) with Eq.(38) in
Ref.~\cite{1} tells us that $(\Delta_n x)^2$ just has an additional
temperature factor $coth({\frac {\beta\hbar\omega}{2}})$.

  Finally, we give two examples of $\rho_s(x,t)$ to end this section. For
$n=0$, we have
\begin{equation}
\rho_s(x,t) = \sqrt{{\frac {1}{2\pi(\Delta_0 x)^2}}}
   exp\{-{\frac {(x-<x>)^2}{2(\Delta_0 x)^2}}  \;.
\end{equation}
Using the relation between the thermalized squeezed state and squeezed
thermalized state in Ref.~\cite{18}(1991), one can find that the last equation
with $t=0$ is consistent with Eq.(6.6a) in Ref.~\cite{20}(1993).
For $n=1$, we have
\begin{equation}
\rho_s(x,t) = \sqrt{{\frac {2}{\pi(\Delta_0 x)^2}}}
   exp\{-{\frac {(x-<x>)^2}{2(\Delta_0 x)^2}}\{{\frac {1}{2}}
   sech^2({\frac {\beta\hbar\omega}{2}})
   [({\frac {(x-<x>)^2}{2(\Delta_0 x)^2}}-{\frac {3}{2}})^2-{\frac {3}{2}}]
        +{\frac {(x-<x>)^2}{2(\Delta_0 x)^2}}\}  \;.
\end{equation}
From the last equation, we see that by introducing a finite temperature effect,
$x$-polynomial factor in the expression of $\rho_s(x,t)$ is not just the
quadratic term of $(x-<x>)$ as Eq.(52) in Ref.~\cite{1}, but have additional
quadruplicate term of $(x-<x>)$. The appearance of the quadruplicate term is
understandable because thermalizing a nonclassical state amounts to doubling
the freedom number of the systems within the framework of thermofield dynamics
\cite{16}.

From Eq.(55) to the last equation, we give some explicit results only about
the thermalized squeezed number state. As for the thermalized displaced number
state, taking $z=0$ in Eqs.(55)---(58), one can obtain the corresponding
results about them, which are also consistent with those in the literature.

\section{Conclusion}
\label{5}

In this paper, we have given the wavefunctions of the thermalized displaced
number and squeezed number states in the coordinate representation.
Furthermore, with the help of the thermal coordinate representation, we obtain
the time-dependent expressions of these wavefunctions. Although the thermal
vacuum wavefunction is time-independent, but either the thermalized displaced
number state or the thermalized squeezed number state varies with time. We also
give the probability densities, average values and variances of the position
coordinate on these stats. Each of the wavefunctions, the time-dependent
wavefunctions, the probability densities, average values or variances here is
consistent with that in the literature when the temperature tends to zero.
Setting $n=0$ in the expressions of this paper, one can obtain results of the
usual thermalized coherent and squeezed states. Additionally, setting $t=0$, one
can obtain the probability densities without considering the time evolution of
these states.

In the thermal coordinate representation, the forms of the wavefunctions
Eqs.(14),(26),(35), (43) and (44) resemble their own zero-temperature limits.
Of course, this resemblance does not exist in the coordinate representation at
all, and the probabilty densities are different from those at the
zero-temperature case. From section IV, one has seen that the thermal
coordinate representation simplified greatly the calculations there. Perhaps,
the thermal coordinate representation would simplify other calculations
related to the thermal non-classical states.

Finally, although the thermalized displaced number and
squeezed number states are discussed, the above results of them can give easily
the corresponding ones of other similar states, such as displaced thermalized
number state, squeezed thermalized number state, $etc.$, with a simple
parameter transformation \cite{18}(1991) \cite{20}(1993). Additionally, no
matter how complicate the expressions (51) and (54) are, it is not difficult
to compute them numerically for a given number $n$. We believe that once the
displaced number state and squeezed number state is prepared in laboratories
some day, the results in the present paper will be useful.

\acknowledgments
This project was supported jointly by the President Foundation of
Shanghai Jiao Tong University and the National Natural Science Foundation of
China with grant No. 19875034.

 
\end{document}